\documentclass[12pt]{article}

\usepackage{amssymb,amsmath}

\usepackage{graphicx}
\DeclareGraphicsExtensions{.pdf,.png,.jpg}

 \catcode`\@=11 \@addtoreset{equation}{section}\catcode`\@=12
\newcommand{\R}{ {\mathbb R} }
\newcommand{\fnm}{\footnotemark}
\newcommand{\fnt}{\footnotetext}

\begin{document}

 \begin{center}

 \large \bf On non-exponential cosmological  solutions with two factor spaces
of dimensions  $m$ and   $ 1$ in the Einstein-Gauss-Bonnet model  with a $\Lambda$-term 
  \end{center}

 \vspace{0.3truecm}

 \begin{center}

  K. K. Ernazarov

\vspace{0.3truecm}

 \it Institute of Gravitation and Cosmology,
 RUDN University, 6 Miklukho-Maklaya ul.,
 Moscow 117198, Russia

 \end{center}

\begin{abstract}

We consider a $(m+2)$-dimensional  Einstein-Gauss-Bonnet model with the cosmological $\Lambda$-term.
We restrict the metrics to be diagonal  ones and find for certain 
$\Lambda = \Lambda(m)$ class of cosmological solutions with  non-exponential time dependence 
of two scale factors of dimensions  $m > 2$ and $1$. 
Any solutions from this class describes an accelerated expansion of $m$-dimensional subspace 
and tends asymptotically to isotropic solution  
with exponental dependence of scale factors.

\end{abstract}

%  {\bf Keywords:} Gauss-Bonnet,  variation of G, accelerated expansion of the Universe

\section{Introduction}

In this paper we consider a $D$-dimensional gravitational model
with Gauss-Bonnet term and cosmological term $\Lambda$, which
extend the model with $\Lambda =0$ from ref. \cite{Deruelle}.

Earlier a wide class of exact cosmological solutions with  diagonal metrics in this model 
with two and three different Hubble-like parameters, e.g.
stable ones   with exponential behaviour of scalar factors, 
were studied in numerous publications, see \cite{ElMakObOsFil}-\cite{ChTop-17}.
The main part of these solutions have constant Hubble-like parameters, some of solutions
deal with the case $\Lambda =0$.  

Recently in ref. \cite{Pavl-16} a systematic study of dynamical solutions in 
low dimensions with two non-constant Hubble-like parameters $H(t)$ and $h(t)$, corresponding
to $3$- and $l$-dimensional factor spaces  was
started for $ l = 1, 2$ ($l$  is dimension of the internal space) and arbitrary $\Lambda$.  We note that in ref. \cite{Pavl-17} there is a continuation of this study for the case of general  $l$.
   
Here we deal with a class of exact dynamical solutions with  non-constant $H(t)$ and $h(t)$,
which correspond to $m$-dimensional ($m > 2$ ) and one-dimensional factor spaces,
when some fine-tuned $\Lambda =\Lambda (m)$ is chosen.
The solution describes an accelerated expansion of $m$-dimensional subspace.  
 
The structure of the paper is as follows. In Section 2 we present a setup. A class of exact cosmological solutions with   diagonal metrics is found  for certain $\Lambda$ in Section 3. In Section 4 we consider examples
of solutions for $m = 3,4,5$.

\section{The set up}

The action of the model reads
\begin{equation}
  S =  \int_{M} d^{D}z \sqrt{|g|} \{ \alpha_1 (R[g] - 2 \Lambda) +
              \alpha_2 {\cal L}_2[g] \},
 \label{2.0}
\end{equation}
where $g = g_{MN} dz^{M} \otimes dz^{N}$ is the metric defined on
the manifold $M$, ${\dim M} = D$, $|g| = |\det (g_{MN})|$, $\Lambda$ is
the cosmological term, $R[g]$ is scalar curvature,
$${\cal L}_2[g] = R_{MNPQ} R^{MNPQ} - 4 R_{MN} R^{MN} +R^2$$
is the standard Gauss-Bonnet term and  $\alpha_1$, $\alpha_2$ are
nonzero constants.

We consider the manifold
\begin{equation}
   M = \R  \times   M_1 \times \ldots \times M_n 
   \label{2.1}
\end{equation}
with the metric
\begin{equation}
 g= - dt \otimes dt + \sum_{i=1}^{n} e^{2\beta^i(t)}  dy^i \otimes dy^i.
 \label{2.2}
\end{equation}
We have  the  set of  equations \cite{ErIvKob-16} 
\begin{eqnarray}
     E = G_{ij} h^i h^j + 2 \Lambda  - \alpha G_{ijkl} h^i h^j h^k h^l = 0,
         \label{2.3} \\
         Y_i =  \frac{d L_i}{dt}  +  (\sum_{j=1}^n h^j) L_i -
                 \frac{2}{3} (G_{sj} h^s h^j -  4 \Lambda) = 0,
                     \label{2.4}
          \end{eqnarray}
where $h^i = \dot{\beta}^i$,           
 \begin{equation}
  L_i = L_i(h) = 2  G_{ij} h^j
       - \frac{4}{3} \alpha  G_{ijkl}  h^j h^k h^l  
       \label{2.5},
 \end{equation}
 $i = 1,\ldots, n$, where  $\alpha = \alpha_2/\alpha_1$. Here
\begin{equation}
G_{ij} = \delta_{ij} -1, \qquad   G_{ijkl}  = G_{ij} G_{ik} G_{il} G_{jk} G_{jl} G_{kl}
\label{2.6}
\end{equation}
are, respectively, the components of two  metrics on  $\R^{n}$ \cite{Iv-09,Iv-10}. 
The first one is a 2-metric and the second one is a Finslerian 4-metric.
For $n > 3$ we get a set of forth-order polynomial  equations.

Here we present a class of solutions to the set of equations (\ref{2.3}), 
(\ref{2.4}) of the following form
\begin{equation}
  \label{3.1}
   (h^i(t))  =(\overbrace{H(t), \ldots, H(t)}^{m},  h(t)).
\end{equation}
where $H(t)$ is the Hubble-like parameter corresponding  
to  $m$-dimensional factor space with $m > 2$, and  $h(t)$ is the Hubble-like parameter 
corresponding to  one-dimensional factor space.

We put 
\begin{equation}
  \label{3.2}
   H(t) > 0,   \qquad  H^2 + \dot{H} > 0,   
\end{equation}
where $t > 0$,  for a possible description of an  accelerated expansion of a
$3$-dimensional subspace (which may describe our Universe)  in our epoch 
($t_0 > 0$ \cite{Riess,Perl}.
 
Here we put $\alpha < 0$.\fnm[1]\fnt[1]{There are certain evidences which favor $\alpha > 0$,  
e.g.,  argument coming from  string theory; see also ref. \cite{Pavl-17}. Here 
as in ref. \cite{Pavl-17} we do not restrict ourselves by string theory (induced) models and other ones which demand $\alpha > 0$.}
The dynamical equation (\ref{2.4}) for $H$ has the following form
 \begin{equation}\label{3.3}
\begin{gathered}
 3(m-1)\dot{H}+3\dot{h}+2\alpha(m-1)(m-2)(2((m-3)H+3h)H\dot{H}+((m-3)\dot{H}+3\dot{h})H^2)\\
+2\alpha m(m-1)(m-2)(m-3)H^4+2\alpha(m-1)(m-2)(4m-3)hH^3\\
+2(m-1)((1+3\alpha h^2)m-6\alpha H^2)H^2+(4m-3)hH+3h^2=4\Lambda
\end{gathered}
\end{equation}
The dynamical equation corresponding to $h$   reads as follows
 \begin{equation}
\begin{gathered}
\label{3.4}
3m\dot{H}+6\alpha m(m-1)(m-2)H^2\dot{H}+2\alpha m^2(m-1)(m-2)H^4\\
+2\alpha m(m-1)(m-2)hH^3+m(2m+1)H^2+mhH=4\Lambda
\end{gathered}
\end{equation}
The equation (\ref{2.3}) may be written the following form:
 \begin{equation}\label{3.5}
\begin{gathered}
   \alpha m(m-1)(m-2)(m-3)H^4+4\alpha m(m-1)(m-2)hH^3 \\
+m(m-1)H^2+2mhH=2\Lambda
\end{gathered}
\end{equation}
The relation (\ref{3.5}) is solvable with respect to h and we obtain the following expression
\begin{equation}
\label{3.6}
 h=-\frac{1}{2}\frac{\alpha m(m-1)(m-2)(m-3)H^4+m(m-1)H^2-2\Lambda}{m\Biggl(2\alpha (m-1)(m-2)H^2 +1\Biggr)H}
\end{equation}
and then substituting this value of h into Eqs. (\ref{3.3}) and (\ref{3.4}).
 We obtain the relation for $ \dot{ H}$:
\begin{equation}
\label{3.7}
\dot{H}=-\frac{1}{2}\frac{\alpha m(m-1)(m-2)(m+1)H^4+m(m+1)H^2-2\Lambda}{m\Biggl(2\alpha (m-1)(m-2)H^2 +1\Biggr)}
\end{equation}
and from the formula (\ref{3.3}) it is possible to obtain the relation for $\dot{ h}$ :
\begin{equation}\label{3.8}
\begin{gathered}
\dot{h}=\frac{1}{4}\frac{1}{m^2(2\alpha(m-1)(m-2)H^2+1)^3H^2}\times\\
\times \Biggl[2\alpha^3m^2(m+1)(m-3)(m-1)^3(m-2)^3H^{10}\\
+\alpha^2m^2(m+1)(3m-13)(m-1)^2(m-2)^2H^8\\
+2\alpha m(m-1)(m-2)\Biggl((1+4\alpha\Lambda)m^3-3m^2-4(1+7\alpha\Lambda)m+24\alpha\Lambda\Biggr)H^6\\
+m(m-1)\Biggl((1+12\alpha\Lambda)m^2+(1+4\alpha\Lambda)m-56\alpha\Lambda\Biggr)H^4\\
-4\Lambda\Biggl(6\alpha\Lambda m^2 -(1+18\alpha\Lambda)m+12\alpha\Lambda\Biggr)H^2-4\Lambda^2\Biggr].
\end{gathered}
\end{equation}

\section{Exact solutions}

Now we consider special solution to equations from the previous section. 
As it can be seen from the expressions (\ref{3.7}) and (\ref{3.8}),
 the problem of finding a general integral in an analytical form from the above expressions causes enormous difficulties, 
i.e. the total integral can be calculated only by numerical methods with varying degrees of accuracy. 
But in the case when $ \alpha = -1$ 
\begin{equation}\label{3.9}
 \Lambda = \frac{1}{8}\frac{m(m+1)}{(m-1)(m-2)}
\end{equation}
these expressions are greatly simplified and, by calculating the integrals, one can find such dynamic exact solutions as $ H (t)$ and $ h (t)$. We already know that in this value of $\Lambda$ we obtain a solution with zero variation of $G$  \cite{ErIv-17}.
By substituting these values of $ \alpha$ and $\Lambda$ into  (\ref{3.7}) and (\ref{3.8}) we obtain 
\begin{equation}\label{3.10}
 \dot{H}=\frac{1}{8}\frac{(m+1)\Biggl(1-2(m-1)(m-2)H^2\Biggr)}{(m-1)(m-2)}
\end{equation}
and
\begin{eqnarray}
\label{3.11}
 \dot{h}=\frac{(m+1)}{64} \times  \qquad  \qquad \qquad  \\ \nonumber 
  \frac{\Biggl(4(m-3)(m-1)^2(m-2)^2H^4+8(m-1)(m-2)H^2-(m+1)\Biggr)}{(m-1)^2(m-2)^2H^2}
\end{eqnarray}

Then, we calculate the integral of (\ref{3.10}) with the help of a tabular integral and taking 
into account the initial condition $H (0) = 0 $, we find the following formulas for H for arbitrary m.
\begin{equation}\label{3.12}
 H=\frac{4b}{m+1}th(b t)
\end{equation}
where
\begin{equation}\label{3.13}
b=\frac{m+1}{4\sqrt{2(m-1)(m-2)}}
\end{equation}
Now we find the expression $ \frac{dh}{dH}$ with the help of equations (\ref{3.10}) and (\ref{3.11}):
\begin{equation}\label{3.14}
\frac{dh}{dH}=-\frac{1}{8}\Biggl(2(m-3)+\frac{(m+1)}{(m-1)(m-2)H^2}\Biggr)
\end{equation}
From the last equation, we compute the integral and taking into account the initial condition $H (0) = 0 $, 
we find the following formulas for h for arbitrary m.
\begin{equation}\label{3.15}
 h=\frac{1}{8}\Biggl(\frac{(m+1)}{(m-1)(m-2)H}-2(m-3)H\Biggr)
\end{equation},
i.e.
\begin{equation}\label{3.16}
 h = b\Biggl(cth(b t)- \frac{m-3}{m+1}th(b t)\Biggr)
\end{equation}

As it can be seen from the formulas (\ref{3.12}) and (\ref{3.16}), as $ t \to \infty $, the parameters 
$H(t)$ and $h(t)$ approach to the same value 
$ H_{st} = (2(m-1)(m-2))^{-\frac{1}{2}}$ , i.e., anisotropic  solution  for Hubble-like parameters turns to static isotropic solution from ref. \cite{Ivas-16} with critical value of $\Lambda$.

{\bf Accelerated expansion.}

The differencial equations $\frac{d\beta^i}{dt}=h^i$, $i = 1, \dots, m +1$ could be readily integrated.
We get for $t > 0$

\begin{equation}
\label{B.3}
%\begin{gathered}
 \beta^i(t)=\frac{4}{m+1}\ln (ch(b t)) + \beta^i_0,
%\end{gathered}
\end{equation}
where $\beta^i_0$ are constants, $i=1, \dots,m$, and 

\begin{equation}
\label{B.4}
%\begin{gathered}
 \beta^{m+1}(t)= \ln (sh(b t))-\frac{m-3}{m+1} \ln (ch(b t)) + \beta^{m+1}_0,
%\end{gathered}
\end{equation}
and $\beta^{m+1}_0$ is constant

For scale factors we find 
\begin{equation}
\label{B.5}
%\begin{gathered}
a_i\left(t\right)=e^{\beta^i}=a_{i0}\Biggl(ch\left(b t \right)\Biggr)^{\frac{4}{m+1}}, 
%\end{gathered}
\end{equation}
$i=1, \dots, m$ and

\begin{equation}\label{B.6}
%\begin{gathered}
a_{m+1}\left(t\right)=e^{\beta^{m+1}}=
a_{m+1,0} (sh\left(b t \right) \Biggl(ch\left(b t \right)\Biggr)^{\frac{3-m}{m+1}}.  
%\end{gathered}
\end{equation}

The positive constants $a_{i0} = e^{\beta^{i}_0}$ and $a_{m+1,0} = e^{\beta^{m+1}_0}$
may be restricted by isotropy condition for our $3d$ Universe:
 $a_{i0} = a_{0}$, $i =1,2,3$. 

We obtain
\begin{equation}
\label{B.7}
%\begin{gathered}
\frac{da_i\left(t\right)}{d t}=
a_{0} \frac{4b}{m+1}\Biggl(ch\left(b t \right)\Biggr)^{\frac{3-m}{m+1}}sh(b t), 
%\end{gathered}
\end{equation}
and 
\begin{equation}\label{B.8}
%\begin{gathered}
\frac{d^2}{d^2t}a_i\left( t \right)=a_{0}\frac{4b^2}{m+1} \Biggl(ch\left(b t \right)\Biggr)^{\frac{2(1-m)}{m+1}}
\Biggl[ ch^2(b t) + \frac{3-m}{m+1}sh^2\left(b t \right)\Biggr], 
%\end{gathered}
\end{equation}
$i=1,2,3$.

It may be readily verified  that the inequality 
$d^2 a_i /d^2 t > 0$ is  obeyed for all values of $t >0$.

{\bf Non-singular begaviour.}
We note that in the limit $t \to + 0$ we get $a_i(t) \to a_{i0}$, 
$i=1, \dots, m$, and $a_{m+1} \sim t^2$, that is a Milne-type behaviour of 
scalar factors. It seems that our solution is a non-singular one with a horizon
at $t = +0$ and a so-called ``black universe'' under horizon \cite{BDM}. 
But this may be a subject of a separate research.

\section{Examples}

Here we consider three special solutions for $m = 3,4,5.$ 

\subsection{The case $m = 3$}

Let us consider the case $ m = 3$, which is a special one 
since in this case the internal space is isotropic.
We get
\begin{equation}
\Lambda= \frac{3}{4}
\end{equation}
and from (\ref{3.12}) and (\ref{3.16}) we obtain
\begin{equation}
H=\frac{1}{2}th(t), \qquad h=\frac{1}{2}cth(t)
\end{equation}
and $b = 1$.

The graphical representation of the functions $H$ and $h$ are shown in Fig. 1.

\begin{figure}[h]
\center{\includegraphics[scale=0.35]{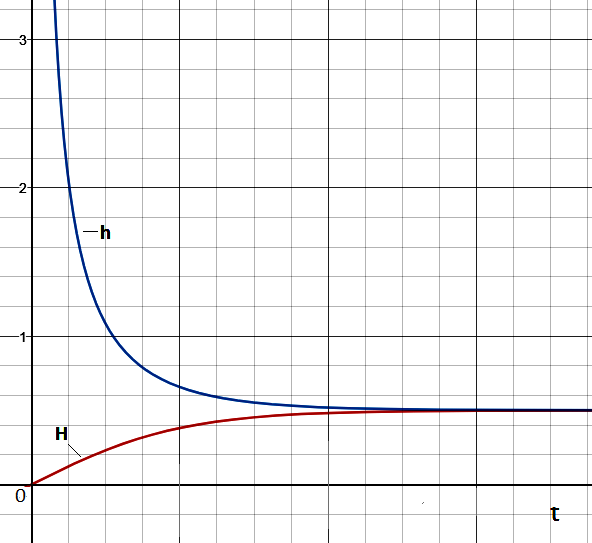}}
\caption{Time evolution of the Hubble-like parameters $H$ and $h$ in the case $m=3$ and $\Lambda=\frac{3}{4}$}
\label{Fig 1}
\end{figure}

Here our solution has an attractor (isotropic) static solution for $t \to + \infty$, i.e.
 $H(t) \to H_{st}=\frac{1}{2}$ and $h(t) \to h_{st}=\frac{1}{2}$.

\subsection{The case $ m = 4$}

Now we consider the case $ m = 4$. We obtain 
\begin{equation}
\label{3.21}
 \Lambda=  \frac{5}{12},
\end{equation}
\begin{equation}
\label{3.22}
H=\frac{1}{\sqrt{12}}th(bt)
\end{equation}
and
\begin{equation}
\label{3.23}
h=\frac{5}{4\sqrt{12}}\Biggl(cth(bt)-\frac{1}{5}th(bt)\Biggr),
\end{equation}
where $b=\frac{5}{4\sqrt{12}}$.

\subsection{The case $ m = 5$}

In the case $ m = 5$ we get
\begin{equation}\label{3.26}
 \Lambda= \frac{5}{16},
\end{equation},
\begin{equation}\label{3.27}
H=\frac{1}{\sqrt{24}}th(bt)
\end{equation}
and
\begin{equation}\label{3.28}
h=\frac{1}{2\sqrt{24}}\Biggl(3cth(bt)-th(bt)\Biggr),
\end{equation}
where $b=\frac{3}{2\sqrt{24}}$.

Here we get  $H(t) \to H_{st}=\frac{1}{\sqrt{24}}$ and $h(t) \to h_{st}=\frac{1}{\sqrt{24}}$,
as  $t \to + \infty$.

\section{Conclusions}

We have considered the  $D$-dimensional  Einstein-Gauss-Bonnet (EGB) model
with the $\Lambda$-term and two constants $\alpha_1$ and $\alpha_2$.  
By using the  ansatz with diagonal  cosmological  metrics, we have found 
for $D =  m + 2 $, $\alpha = \alpha_2 / \alpha_1 = -1$ and certain  $\Lambda = \Lambda(m)$   
 a class of solutions with  non-exponential 
time dependence of $2$ scale factors, governed by two Hubble-like parameters $H(t) >0$,  
$h(t)$, corresponding to submanifolds of  
dimensions $m > 2$ and  $1$, respectively.

For any $m >2$ the Hubble-like parameters have  asymptotically 
 isotropic limits $H(t) \to H_{st} >0$, $h(t) \to h_{st} = H_{st}$, 
which corresponds to critical isotropic exponential solutions from ref. \cite{Ivas-16}.
For any  $m >2$ we have obtained an accelerated expansion of $m$-dimensional subspace.

%\begin{center}
 {\bf Acknowledgments}
%\end{center}
This work was supported in part by the Russian Foundation for
   Basic Research grant No. 16-02-00602 and by the Ministry of Education of the Russian Federation 
   (the  Agreement number 02.a03.21.0008 of 24 June 2016).

%\newpage


\small

\begin{thebibliography}{99}



 \bibitem{Deruelle}
 N. Deruelle, On the approach to the cosmological
 singularity in quadratic theories of gravity: the Kasner
 regimes,   Nucl. Phys. B  {\bf  327},  253-266 (1989).


\bibitem{ElMakObOsFil}
 E. Elizalde, A.N. Makarenko, V.V. Obukhov, K.E. Osetrin  and
 A.E. Filippov,  Stationary vs. singular points in an accelerating
 FRW cosmology derived from six-dimensional Einstein-Gauss-Bonnet
 gravity,  Phys. Lett. B {\bf  644}, 1-6 (2007);  hep-th/0611213.

  
\bibitem{Pavl}
 S.A. Pavluchenko,
 On the general features of Bianchi-I cosmological models in
 Lovelock gravity,  Phys. Rev. D {\bf  80}, 107501 (2009);
 arXiv: 0906.0141.


\bibitem{Iv-09}
  V.D. Ivashchuk,
  On anisotropic Gauss-Bonnet cosmologies in (n + 1) dimensions,
  governed by an n-dimensional Finslerian 4-metric,  Grav. Cosmol.
  {\bf 16}(2), 118-125 (2010); arXiv: 0909.5462.

 \bibitem{Iv-10}
 V.D. Ivashchuk,  On cosmological-type solutions in
 multidimensional model with  Gauss-Bonnet term,
 Int. J. Geom. Meth. Mod. Phys.
 {\bf 7}(5), 797-819 (2010);  arXiv: 0910.3426.

\bibitem{ChPavTop}
D. Chirkov, S. Pavluchenko and A. Toporensky, Exact exponential solutions
in Einstein-Gauss-Bonnet flat anisotropic cosmology,
 Mod. Phys. Lett. A {\bf  29},  1450093 (11 pages) (2014);  arXiv:1401.2962.

\bibitem{ChPavTop1}
D. Chirkov, S.A. Pavluchenko and A. Toporensky,
Non-constant volume exponential solutions in higher-dimensional
Lovelock cosmologies,  Gen. Relativ. Gravit.  {\bf 47}: 137 (33 pages) (2015); arXiv: 1501.04360.

\bibitem{IvKob}
 V.D. Ivashchuk and A.A. Kobtsev,
 On exponential cosmological type solutions in the model
 with Gauss-Bonnet term and variation of gravitational constant,
 Eur. Phys. J.  C  {\bf  75}: 177 (12 pages) (2015);
 arXiv:1503.00860.

 \bibitem{Pavl-15}
 S.A. Pavluchenko, Stability analysis of exponential solutions in Lovelock cosmologies,
 Phys. Rev. D {\bf 92}, 104017 (2015); arXiv: 1507.01871.

\bibitem{Pavl-16}
 S.A. Pavluchenko, Cosmological dynamics of spatially flat Einstein-Gauss-Bonnet models in various dimensions: Low-dimensional 
 $\Lambda$-term case, Phys. Rev. D {\bf 94}, 084019 (2016); arXiv: 1607.07347.  
 
 \bibitem{Pavl-17}
 S.A. Pavluchenko,  Cosmological dynamics of spatially 
 flat Einstein-Gauss-Bonnet models in various dimensions: High-dimensional $\Lambda$-term case,
 Eur. Phys. J. C {\bf 77}, 503 (2017); arXiv: 1705.02578.
    
\bibitem{ErIvKob-16}
 K.K. Ernazarov, V.D. Ivashchuk and A.A. Kobtsev,
 On exponential solutions in the Einstein-Gauss-Bonnet cosmology,
 stability and variation of G,  
   Grav.  Cosmol., {\bf 22} (3), 245-250 (2016).

\bibitem{Ivas-16}
V.D. Ivashchuk, On stability of exponential cosmological solutions
with non-static volume factor in the Einstein-Gauss-Bonnet model,
 Eur. Phys. J.  C {\bf 76} 431 (2016); arXiv: 1607.01244v2.

\bibitem{Ivas-16-2}
V.D. Ivashchuk,  On Stable Exponential Solutions in Einstein-Gauss-Bonnet 
Cosmology with Zero Variation of G,  Grav. Cosmol., {\bf 22} (4), 329-332 (2016); 
see corrected version in arXiv: 1612.07178.

\bibitem{ErIv-17}
 K.K. Ernazarov and V.D. Ivashchuk,
 Stable exponential cosmological solutions with zero variation of G
 in the Einstein-Gauss-Bonnet model with a $\Lambda$-term,
 Eur. Phys. J.  C {\bf 77} 89 (2017); arXiv: 1612.08451. 

\bibitem{ErIv-17-2}
 K.K. Ernazarov, V.D. Ivashchuk, Stable exponential cosmological solutions with zero variation of G and three
 different Hubble-like parameters in the Einstein-Gauss-Bonnet model with a $\Lambda$-term, 
 Eur. Phys. J. C (2017) 77:
 402 (7 pages); arXiv:1705.05456.

\bibitem{ChTop-17}
D.M. Chirkov and A.V. Toporensky,
On stable exponential cosmological solutions in the EGB model with a $\Lambda$-term in
dimensions $D = 5, 6, 7, 8$; arXiv:1706.08889.


\bibitem{Riess}
 A.G. Riess  et al. Observational evidence from supernovae
 for an accelerating  universe and a cosmological constant,
  Astron. J. {\bf 116}, 1009-1038 (1998).

\bibitem{Perl}
 S. Perlmutter  et al. Measurements of Omega and Lambda from 42 High-Redshift  Supernovae.
  Astrophys.  J. {\bf 517},  565-586 (1999).


\bibitem{BDM}
K.A. Bronnikov, H. Dehnen, V.N. Melnikov, 
Regular black holes and black universes, Gen. Rel. Grav. 39, 973-987 (2007); gr-qc/0611022.

\end{thebibliography}
\end{document}